\documentclass[epj]{svjour}
\usepackage{graphics}
\usepackage{graphicx,t1enc}
\usepackage{epsfig}
\usepackage{amssymb,amsmath}

\begin{document}
\title{Synchronization of multi-phase oscillators: An Axelrod-inspired model}
\author{Marcelo N. Kuperman and Dami\'an H. Zanette}
\institute{Consejo Nacional de Investigaciones Cient\'{\i}ficas y
T\'ecnicas, Centro At\'omico Bariloche and Instituto Balseiro, 8400
Bariloche, R\'{\i}o Negro, Argentina}
\date{Received: date / Revised version: date}

\authorrunning{M. N. Kuperman, D. H. Zanette}
\titlerunning{Synchronization of multi-phase oscillators}

\abstract{Inspired by Axelrod's model of culture dissemination, we
introduce and analyze a model for a population of coupled
oscillators where different levels of synchronization can be
assimilated to different degrees of cultural organization. The state
of each oscillator is represented by a set of phases, and the
interaction --which occurs between homologous phases-- is weighted
by a decreasing function of the distance between individual states.
Both ordered arrays and random networks are considered. We find that
the transition between synchronization and incoherent behaviour is
mediated by a clustering regime with rich organizational structure,
where some of the phases of a given oscillator can be synchronized
to a certain cluster, while its other phases are synchronized to
different clusters. }
\maketitle

\section{Introduction}

Robert Axelrod's model for the dissemination of culture
\cite{Axelrod} aims at giving a simplified picture of the processes
that shape the distribution of cultural features in a society, in
particular, the balance between the trend to convergence due to
cultural affinity, and the mechanisms that maintain diversity. This
model has attracted the attention of physicists because of its
nontrivial phenomenology, which includes the existence of a variety
of absorbing states, critical phenomena, and non-monotonic dynamics
\cite{castellano}. Among other topics, the transition between
homogeneous and heterogeneous from states under variation of the
number of cultural traits \cite{castell2}, the effects of noise
\cite{klemm1} and non-trivial underlying topologies \cite{klemm2},
as well as the mean-field limit \cite{redner} have been
characterized.

In its original formulation \cite{Axelrod}, Axelrod's model
described the cultural profile of each agent in a population by a
vector of \emph{features}. Each feature can adopt a number of
\emph{traits}. Agents are situated at the nodes of a bidimensional
square array, and interaction occurs between nearest neighbours. Two
neighbour agents interact with a probability proportional to their
cultural overlap, i.e. to number of features in their profiles whose
traits coincide. As the result of the interaction, the profiles
change, approaching each other. If the two profiles exhibit
completely different traits, interaction is impossible. This system
reaches a frozen state where any two neighbours have either
identical or completely different profiles. The final state can be
homogeneous, with all the population having the same profile, or
consist of coexisting domains, thus maintaining cultural diversity
\cite{Axelrod,castellano}. It turns out that, as the number of
possible traits per feature grows, there is a transition from
homogeneity to diversity \cite{castell2}.

From a physicist's viewpoint, there is a strong similarity between
the self-organization process that determines the distribution of
cultural profiles in Axelrod's model, and the evolution of a
spatially extended system able to develop either homogeneous states
or phase coexistence. Synchronization of distributed interacting
oscillators is an example \cite{ermen,saka,manru}. In these systems,
the attractive interactions that may lead to the formation of
synchronized domains plays a role similar to the tendency of
interacting individuals to become culturally more similar in
Axelrod's model. On the other hand, diversity between individual
attributes of oscillators impedes their synchronization, like too
much cultural divergence does not allow interaction.

Inspired by Axelrod's model, in this paper we introduce and analyze
a system of interacting oscillators, where the state of each
oscillator is characterized by a set of several phases. Each phase
would correspond, in Axelrod's model, to a cultural feature.
Homologous phases of different oscillators are coupled to each
other, with a coupling intensity that depends on the overall
distance between the states of the two oscillators. We show that, in
this system, the synchronization transition can be straightforwardly
identified with Axelrod's transition from cultural homogeneity to
diversity. Additionally, we disclose an intermediate regime with
rich organizational structure, where clusters of mutually
synchronized oscillators, different for each phase, spontaneously
appear.

\section{Model}

We consider a population of $N$ oscillators, each of them occupying
a node in a network. The state of oscillator $i$ is characterized by
a set of $F$ phases, $\phi_i^f(t) \in (0,2\pi)$ ($f=1,\dots , F$).
The dynamics of the phases $\phi_i^f(t)$ is governed by
Kuramoto-like equations \cite{Kuramoto,manru}
\begin{equation} \label{eq1}
\dot \phi_i^f = \frac{1}{\nu_i} \sum_{j \in {\cal N}_i} k_{ij} \sin
(\phi_j^f-\phi_i^f),
\end{equation}
where the sum runs over the oscillators $j$ connected to $i$ by
network links, which define the neighbourhood ${\cal N}_i$ of $i$.
The number of oscillators in ${\cal N}_i$ is $\nu_i$. The
non-negative coupling constant $k_{ij}$ is computed as a prescribed
function of the distance $D_{ij}$ between $i$ and $j$,
\begin{equation} \label{kij}
 k_{ij} = K(D_{ij}).
\end{equation}
The distance $D_{ij}$, in turn, characterizes the difference between
the individual states of oscillators $i$ and $j$, as
\begin{equation} \label{Dij}
D_{ij} = \frac{1}{2F} \sum_{f=1}^F \left[ 1-\cos
(\phi_j^f-\phi_i^f)\right] .
\end{equation}
Note that $D_{ij}$ is non-negative, and $D_{ij}=0$ if and only if
$\phi_j^f=\phi_i^f$ for all $f=1,\dots , F$. Moreover, the maximal
possible value of the distance is $D_{ij}=1$. For the function
$K(D)$, which defines the coupling constant through Eq. (\ref{kij}),
we choose
\begin{equation} \label{K(D)}
K(D) = \left\{
\begin{array}{ll}
1-(\alpha D)^r & \mbox{for $D< \alpha^{-1}$,} \\ \\
0 & \mbox{otherwise},
\end{array}
\right.
\end{equation}
with $\alpha >1$ and $r>0$. Thus, the coupling constant is maximal,
$k_{ij}=1$, when the distance $D_{ij}$ vanishes, and decreases
monotonically as $D_{ij}$ grows. It reaches zero for $D_{ij}=
\alpha^{-1}$, and $k_{ij}=0$ for larger distances. The exponent $r$
controls the behaviour of $k_{ij}$ at small distances: for $r<1$ and
$r>1$, the  coupling constant at $D_{ij}=0$  displays, respectively,
a cusp and a flat maximum. For $r=1$, the dependence of $k_{ij}$ on
the distance is linear.

When the coupling constant $k_{ij}$ is positive, the summand in the
right-hand side of Eq. (\ref{eq1}) represents an attractive
interaction between $\phi_i^f$ and $\phi_j^f$. Under its action, the
two phases will tend to become mutually synchronized. While the
functional dependence of the interaction couples phases separately
for each value of $f=1, \dots , F$, its intensity --determined by
the coupling constant-- is given by the joint contribution of the
difference of all the $F$ phases between the two oscillators through
their distance $D_{ij}$, Eq. (\ref{Dij}).

The multi-phase oscillator system defined by Eqs. (\ref{eq1}) to
(\ref{K(D)}) can be qualitatively compared to Axelrod's model as
follows. The $F$ phases of each oscillator are associated with an
agent's cultural features. Different values of each phase --which,
in our system, vary continuously over $(0,2\pi)$-- correspond to the
traits of Axelrod's model. The distance $D_{ij}$ gives a measure of
the cultural dissimilarity between two agents. If it is too large,
the two agents do not interact. For small distances, on the other
hand, the intensity --or, in Axelrod's model, the probability-- of
the interaction grows as the distance decreases. Synchronization
between phases of two or more oscillators can be assimilated to
their cultural consensus. This may occur with respect to all the
phases --i.e., all the cultural features-- or just some of them. A
given oscillator could have some of the phases synchronized with a
certain part of the population, and the other phases synchronized
with another part, thus giving rise to a state of partial
synchronization equivalent to a rich diversity of cultural domains.
As in other variants of the model \cite{castellano}, we are here
considering that the structure of the population is defined by a
generic network, instead of the bidimensional array originally
considered by Axelrod.

In our system, we expect that collective synchronization is easier
to achieve when the interaction range with respect to the distance
$D_{ij}$ between the oscillators' states, given by $\alpha^{-1}$ in
Eq. (\ref{K(D)}), is larger. This corresponds, in Axelrod's picture,
to a small number of traits per cultural feature. It has been shown,
in fact, that Axelrod's model exhibits a transition between cultural
homogeneity and diversity as the number of traits in each feature
grows \cite{castell2,castellano}. Similarly, in Eq. (\ref{eq1}) a
synchronization transition is expected to occur depending on the
interaction range $\alpha^{-1}$ being larger or smaller than the
typical distance between the states of any two oscillators in the
population.

Assume that the system is in a fully unsynchronized state, with all
the $F$ phases of all oscillators uniformly distributed over
$(0,2\pi)$. The expected average value for the distance $D_{ij}$
between any two oscillators is
\begin{eqnarray}
\langle D_{ij} \rangle &=& \frac{1}{8\pi^2F}  \sum_{f=1}^F
\int_0^{2\pi}  d \phi_i^f \int_0^{2\pi} d \phi_j^f \left[ 1-\cos
(\phi_j^f-\phi_i^f)\right]\nonumber \\  &=& \frac{1}{2}.
\end{eqnarray}
If this mean distance is larger than $\alpha^{-1}$, the
unsynchronized state should be stable as oscillator pairs do not
interact on the average. Consequently, we predict a transition from
synchronization to desynchronization at the critical point
\begin{equation} \label{alfac}
\alpha_c = 2.
\end{equation}
The width of this transition should be controlled by the dispersion
in the values of $D_{ij}$. Under the same assumptions as for the
calculation of $\langle D_{ij} \rangle$, the mean square dispersion
of the distance turns out to be
\begin{equation} \label{sigmaD}
\sigma_D = \left[ \langle D_{ij}^2 \rangle-\langle D_{ij} \rangle^2
\right]^{1/2}=\frac{1}{\sqrt{8F}}.
\end{equation}
According to this estimation, thus, the transition width should
decrease as the number of phases $F$ grows. On the other hand, since
our estimation is based on the evaluation of the average distance
between just two oscillators, the width does not depend on the
population size $N$.

In the following section, we present numerical results for our
model, Eqs. (\ref{eq1}) to (\ref{K(D)}), for populations distributed
over ordered and random networks. We confirm the above predictions
on the synchronization transition, and show that around the critical
point the population segregates into clusters of mutually
synchronized oscillators, displaying a high degree of organizational
diversity.

\section{Numerical results}

The results presented in this section are the outcome of numerical
calculations made on populations of various sizes, ranging from
$N=100$ to $1000$ oscillators. As for the underlying interaction
network, we consider a one-di\-men\-sion\-al ordered array and
random Erd{\H o}s-R\'enyi networks \cite{redes}. In the former, each
node is connected to its four nearest neighbours, two to each side.
In the latter, the average number of neighbours per node equals
four.

Bearing in mind the analogy with Axelrod's model, we focus our
attention on the organization of the population into synchronized
clusters at asymptotically long times. A synchronized cluster is a
group of oscillators whose phases coincide. Given that we deal with
multi-phase oscillators, where the state of each of them is
described by $F$ phases, we analyze the organization into groups for
each phase separately. Numerically, we consider that two phases are
synchronized if they differ in less than $\Delta \phi= 10^{-4}$. To
statistically characterize the organization into groups we measure
the normalized mean number of groups $g$ and the dispersion
$\sigma_g$, averaging over phases and realizations. The
normalization of the number of groups is performed with respect to
its maximum value, $N$.

A more detailed characterization of the organization into groups is
achieved by introduced a normalized Hamming distance, $h$, averaged
over phases and oscillator pairs, as follows. First, we represent
the synchronization state of each oscillator pair by defining a set
of $F$ matrices $M^f$ ($f=1,\dots,F$), $N \times N$ in size, with
elements
\begin{equation}
M^f_{ij}=\left \{ \begin{array}{ll} 1 & \mbox{\ \
if $\phi^f_i$ and $\phi^f_j$ are  synchronized,}\\
&\\
0 & \mbox{\ \ otherwise}.
\end{array}
\right. \label{ham}
\end{equation}
The Hamming distance between two of these matrices, $M^f$ and
$M^{f'}$, is defined as
\begin{equation}
H_{ff'} = \sum_{i,j=1}^N | M^f_{ij}-M^{f'}_{ij}|.
\end{equation}
It equals zero if and only if the two matrices are identical, and
its maximum value is $N(N-1)$. The normalized Hamming distance $h$
is obtained by averaging $H_{ff'}$ over all the matrix pairs and
normalizing to the maximum value.

\begin{figure}
\includegraphics[width=9cm]{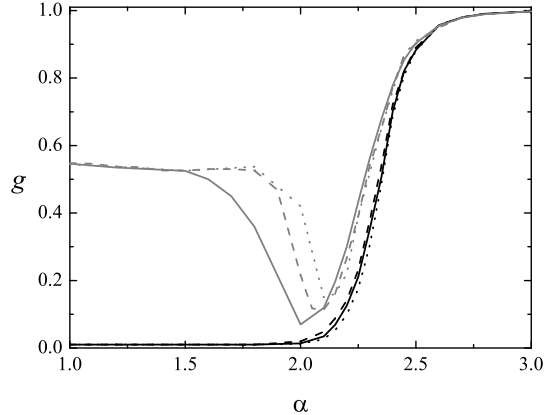}
\caption{Normalized number of groups as a function of $\alpha$ for
the one-dimensional array (grey) and for random networks (black)
with $N=100$ and $F=50$. Each set of curves corresponds to three
values of the exponent $r= 0.1$ (solid), $1$ (dashed), and $3$
(dotted). } \label{ovsd}
\end{figure}

Figure \ref{ovsd} shows results for the normalized number of groups
$g$ as a function of the parameter $\alpha$ of Eq. (\ref{K(D)}), for
various values of the exponent $r$ and both network topologies. A
feature common to all curves is the rather sharp transition from
small  to large values of $g$ as $\alpha$ grows, just above the
critical value predicted in the preceding section, Eq.
(\ref{alfac}). This corresponds, as expected, to a desynchronization
transition as the range of coupling decreases. To the right of the
transition, in fact, the number of groups equals the population
size, reveling a state where all oscillators are mutually
unsynchronized.

Around and above the transition the behaviour is rather independent
of the parameters and of the underlying topology, but a substantial
difference is apparent for $\alpha<2$. While in random networks the
normalized number of groups remains small and close to its minimal
value $g=1/N$, revealing full synchronization of all oscillators in
all their phases, for the one-dimensional array $g$ attains rather
large values, $g  \gtrsim 0.5$. By inspecting single realizations on
the one-dimensional array, we have verified that these large values
of $g$ are due to the occasional occurrence of the so-called
\emph{twisted states} \cite{wsg}. In twisted states, at
asymptotically long times, oscillator phases are not synchronized
but vary linearly along the array. These spatially-correlated
distributions of phases are possible due to the underlying
topological order. From the viewpoint of our approach, however, they
represent fully unsynchronized states where the number of groups
equals the size of the population, $N$.  Therefore, their
contribution makes the average normalized number of groups $g$ grow.

For larger values of $\alpha$, as illustrated by Fig.\ref{ovsd},
differences between results for random networks and the
one-dimensional array, as well as for different values of $r$, are
much less pronounced. Thus, in our analysis for $\alpha > 2$, we
focus attention on random networks and fix $r=1$.

\begin{figure}
\includegraphics[width=9cm]{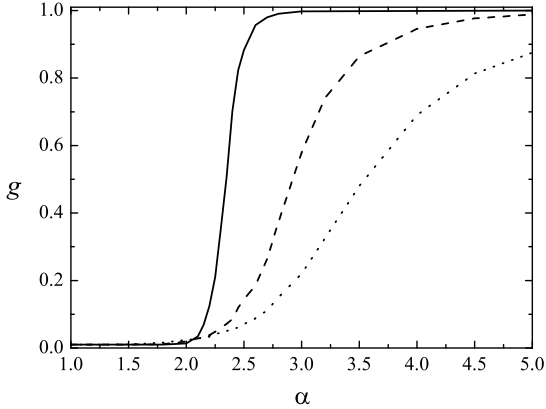}
\caption{Normalized number of groups $g$ as a function of $\alpha$
for $N=100$ and $F=50$ (solid), $10$ (dashed), and $5$ (dotted).}
\label{ga10}
\end{figure}

\begin{figure}
\includegraphics[width=9cm]{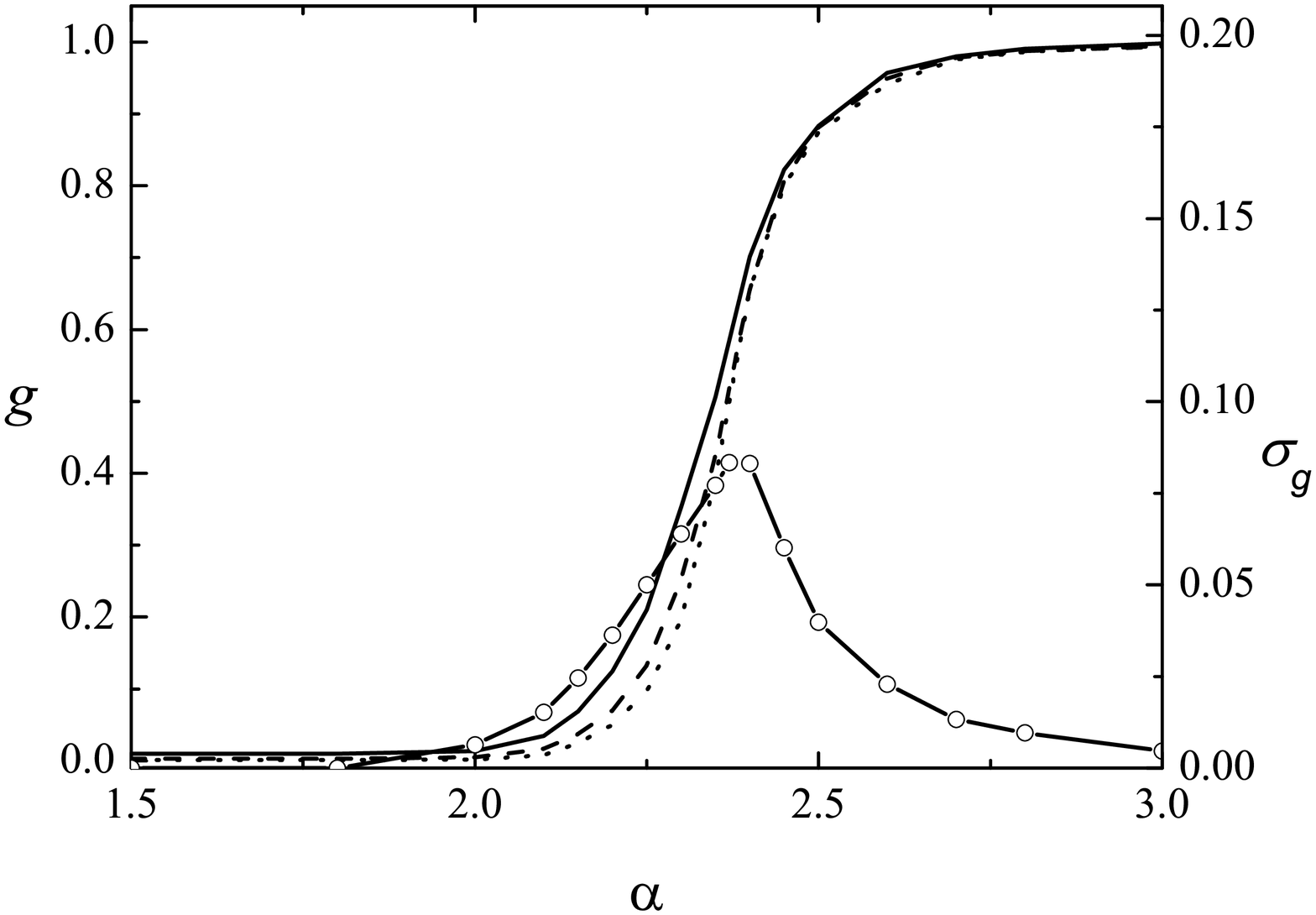}
\caption{Normalized number of groups $g$ as a function of $\alpha$
for $F=50$ and $N=100$ (solid), $N=300$ (dashed), and $N=1000$
(dotted). Empty dots represent the mean square dispersion $\sigma_g$
of the normalized number of groups for $N=100$.} \label{ran}
\end{figure}

We first verify the prediction made in the preceding section, Eq.
(\ref{sigmaD}), that the width of the transition should decrease
with as the number of phases $F$ grows. This is confirmed by the
numerical results shown in Fig. \ref{ga10}, where we plot  the
normalized number of groups $g$ as a function of $\alpha$ for three
values of $F$.
Figure \ref{ran} shows that, in agreement with our prediction, there
is practically no dependence of the transition width on the
population size $N$.

In contrast with the situation for small $\alpha$, the growth of $g$
in the zone of the transition, $\alpha \gtrsim 2$, is due to the
disaggregation of the fully synchronized state into clusters of
mutually synchronized oscillators. Clusters may have various sizes
and, consequently, the number of clusters may vary between
realizations. Also, as we discuss later, oscillators which are
mutually synchronized in some of their phases need not be
synchronized in all of them, so that the structure of clustering in
a given population is not necessarily the same for each phase. We
call this mixed clustering structure {\em cross synchronization}.
The phenomenon of cross synchronization reveals a very rich form of
self-organization, also by comparison with the multi-cultural
configurations of Axelrod's model.

\begin{figure}
\includegraphics[width=9cm]{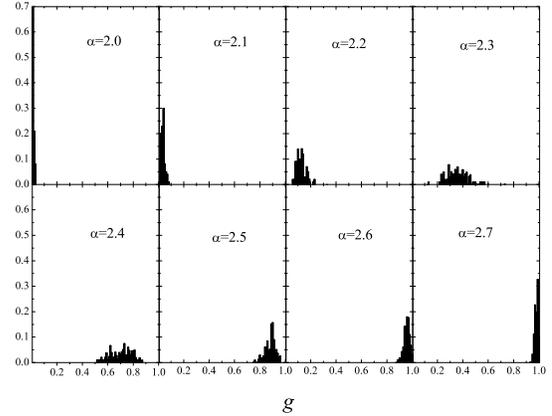}
\caption{Histograms of the normalized number of groups $g$ for
several values of $\alpha$, corresponding to the realizations for
$N=100$ used to calculate the averages and mean square dispersions
shown in Fig. \ref{ran}}. \label{histo}
\end{figure}

Empty dots in Fig. \ref{ran} represent the mean square dispersion,
over phases and realizations, of the normalized number of groups,
for $N=100$ and $F=50$. We recognize the typical fluctuation peak
around the critical point, indicating --in our case-- that the
number of groups is less well-defined in the transition range where
synchronization breaks down. A more detailed picture of the
statistics of groups at the transition is provided by
Fig.\ref{histo}, where we show histograms of the individual values
of the normalized number of groups for several values of $\alpha$.
As $\alpha$ grows, the histogram shifts from a  sharp peak at  $g
\approx 0$ to a sharp peak at $g\approx 1$, passing by broader
distributions of variable width as the transition proceeds.

\begin{figure}
\includegraphics[width=9cm]{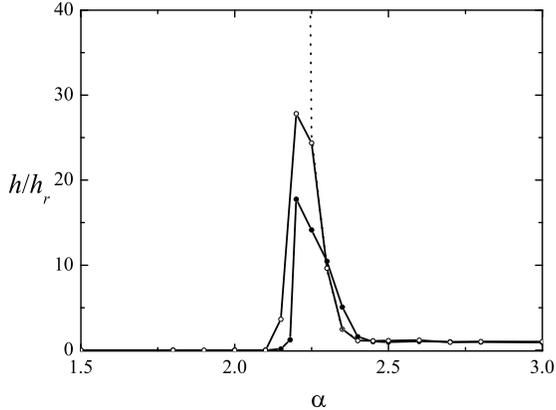}
\caption{Normalized Hamming distance ratio, $h/h_r$, as a function
of $\alpha$, for $N=100$ and $F=50$ on a random network (full dots),
and on the one-dimensional array, either excluding (empty dots) or
including (dotted line) twisted states. } \label{hamdes}
\end{figure}

The phenomenon of cross synchronization addressed to above --i.e.,
the formation of different clustering structures for different
phases-- can be quantified using the normalized Hamming distance
$h$. In fact, if the distribution of oscillators into clusters is
the same in all phases, the matrices $M_{ij}^f$ ($f=1,\dots,F$) are
all identical, and the normalized Hamming distance  vanishes. On the
other hand, different clustering structures for different phases
generally give a positive value of $h$. For a quantitative
evaluation of  cross synchronization, the value of $h$ must be
compared with the normalized Hamming distance $h_r$ for a random
distribution of phases. Numerically, $h_r$ can be computed by
randomly shuffling the phases of all oscillators. Figure
\ref{hamdes} shows, as full dots, the ratio $h/h_r$ as a function of
$\alpha$ for $N=100$ and $F=50$ on a random network. Below the
transition, as expected, $h/h_r=0$. In fact, in the fully
synchronized state only one group --containing all the population--
is formed in all phases. For $\alpha \gg 2$, on the other hand,
$h/h_r \approx 1$, showing that well above the transition the
unsynchronized state is statistically equivalent to a random
distribution of phases. In the intermediate zone, in contrast,
$h/h_r$ exhibits a sharp maximum attaining values close to $20$. It
is therefore in this region that cross synchronization is most
conspicuous.

Cross synchronization occurs also on the one-dimen\-sion\-al array.
In this case, however, its detection through the Hamming distance is
more tricky due to the presence of twisted states below the
transition. In realizations where some of the phases are
synchronized while other are distributed in twisted states the
clustering structures are maximally different and, consequently, $h$
attains very large values. The dotted line in Fig. \ref{hamdes}
represents the ratio $h/h_r$ as a function of $\alpha$ for
realizations on the one-dimensional array. To compare with random
networks, thus, we exclude from the calculation of $h$ the
realizations where twisted states are found. Empty dots in Fig.
\ref{hamdes} stand for the result of this procedure. The phenomenon
of cross synchronization is qualitatively the same as, but stronger
than, in random networks.

\section{Conclusion}

Our study of synchronization in a population of coupled multi-phase
oscillators has been motivated by the affinity between the emergence
of cultural domains in Axelrod's model
\cite{Axelrod,castellano,castell2} and the occurrence of coherent
behaviour in distributed interacting systems
\cite{ermen,saka,manru}. The state of each oscillator is
characterized by a set of phases, which represent the cultural
features of Axelrod's model. We focused our attention on the
long-time asymptotic organization of the oscillators into
synchronized or unsynchronized configurations, under the effect of
pair interactions which are attractive if the individual states are
similar, but are absent if the difference between individual states
grows beyond a given threshold.

Analytic calculations suggest that, as a function of the interaction
threshold, the population undergoes a synchronization transition. If
the threshold is too small, individual states are generally too
different from each other, and the population fails to collectively
synchronize. Conversely, a large threshold enables the attractive
interactions and promotes synchronization. The same calculations
show that the transition becomes sharper if the number of phases per
oscillator grows. On the other hand, the transition width is
independent on the population size.

These analytical predictions were confirmed by numerical
realizations of our system, both on ordered arrays and on random
networks. Numerical results also revealed that, regarding the
organizational structure of the population, the most interesting
regime occurs precisely around the transition. We first found that,
in this region, the synchronized state undergoes disaggregation into
clusters of mutually synchronized oscillators, as observed in other
systems of interacting oscillators \cite{manru}. The number of
clusters and, consequently, their size, varies with the interaction
threshold.

Remarkably, moreover, the segregation of oscillators into clusters
is not necessarily associated with the synchronization of all the
individual phases. In particular, two oscillators may belong to the
same synchronized cluster with respect to some of their phases
--which, in the asymptotic state, adopt identical values-- but to
different clusters with respect to the other phases. This
phenomenon, which we have called cross synchronization, discloses a
highly complex form of collective organization, consisting of
coherent clusters mixed with respect to the internal variables which
specify the individual states. It is interesting to note that
Axelrod's model does not exhibit a regime corresponding to cross
synchronization. This kind of regime would however make sense as a
possible distribution of culture diversity over a real population,
in the form of partially overlapping cultural domains. We have here
presented  a preliminary quantitative analysis of cross
synchronization --aiming, mainly, at giving a means to detect the
phenomenon-- but, clearly, much further work is necessary to fully
characterize its statistical features.

\section*{Acknowledgements}

We acknowledge financial support from CONICET (PIP 5114), ANPCyT
(PICT 04/943), and SECTyP (UNCuyo), Argentina.

\end{document}